# Tumour Control Probability in Cancer Stem Cells Hypothesis

A. Dhawan [1,2], M. Kohandel [1,3], R. P. Hill [4], S. Sivaloganathan [1,3]

[1] Department of Applied Mathematics, University of Waterloo,
Waterloo, Ontario, Canada, N2L 3G1
[2] School of Medicine, Queen's University, Kingston, Ontario, Canada, K7L 3N6
[3] Center for Mathematical Medicine, Fields Institute for Research in Mathematical
Sciences, Toronto, Ontario, Canada, M5T 3J1
4 Department of Medical Biophysics, Princess Margaret Hospital,
Ontario Cancer Institute, University of Toronto,
Toronto, Ontario, Canada, M5G 2M9

**Abstract**

The tumour control probability (TCP) is a formalism derived to compare various treatment regimens of radiation therapy, defined as the probability that given a prescribed dose of radiation, a tumour has been eradicated or controlled. In the traditional view of cancer, all cells share the ability to divide without limit and thus have the potential to generate a malignant tumour. However, an emerging notion is that only a sub-population of cells, the so-called cancer stem cells (CSCs), are responsible for the initiation and maintenance of the tumour. A key implication of the CSC hypothesis is that these cells must be eradicated to achieve cures, thus we define $TCP_S$ as the probability of eradicating CSCs for a given dose of radiation. A cell surface protein expression profile, such as CD44high/CD24low for breast cancer, is often used as a biomarker to monitor CSCs enrichment. However, it is increasingly recognized that not all cells bearing this expression profile are necessarily CSCs, and in particular early generations of progenitor cells may share the same phenotype. Thus, due to the lack of a perfect biomarker for CSCs, we also define a novel measurable $TCP_{CD+}$, that is the probability of eliminating or controlling biomarker positive cells. Based on these definitions, we use stochastic methods and numerical simulations to compare the theoretical $TCP_S$ and the measurable $TCP_{CD+}$. We also use the measurable TCP to compare the effect of various radiation protocols.

**Introduction**

Radiotherapy has become a primary vehicle for cancer therapy, and its continued use as an effective therapeutic and palliative treatment can only be justified if the risks of associated side effects that may be incurred are minimized. Theoretically, this can be posed as an optimization problem, where the risk-to-benefit function for the radiation dosing and scheduling must be optimized. In this work, one key element of this function, the attainable benefit from the treatment, is examined. Classically, the tumour control probability (TCP) has been used as a tool in radiotherapy to measure the probability that the goal of the treatment - the elimination of all clonogenic cells - has been achieved (Munro and Gilbert 1961). Using data in the form of survival fraction curves (which carry information of the proportion of cells that survive a specified dose of radiation), as a probability model for radiation-induced individual clonogenic cell death, the TCP computes the probability of tumour eradication by taking into account factors such as cell proliferation between radiation treatment fractions, and natural cell death rates.

The traditional view of cancer asserts that all cells in a malignant tumour are clonogenic, with genetic and epigenetic differences. An emerging hypothesis is the notion that many cancers are driven by cancer stem cells (CSCs), a subpopulation of cells that have the capacity to proliferate indefinitely and hence to drive and maintain tumour growth. Existence of CSCs has been firmly identified in leukaemia (Bonnet and Dick 1997), and more recently in many solid tumours including breast cancer (Al-Hajj et al. 2003) and brain tumours (Singh et al. 2003, Singh et al. 2004).

Moreover, other works have hypothesized (and demonstrated in the context of hematopoietic cancers) the existence of a hierarchy of cells at various stages of differentiation comprising a tumour, starting with stem cells differentiating into progenitor cells, which differentiate into mature cells. A main implication of the CSC hypothesis is that CSCs can generate all of the cells within a given tumour that lack cancer propagating potential (non-CSCs), and also that the CSCs must be eradicated to control the tumour (Nguyen et al. 2012). Hence, we define $TCP_S$ as the probability of eradicating or controlling CSCs for a given total dose of radiation.

Like normal stem cells, cell surface protein expression profiles are frequently used to identify and isolate CSCs. This includes CD34highCD38low for leukaemia, CD133+ for brain tumours and CD44highCD24low for breast tumours. However, there is growing evidence that not all cells bearing this expression profile are necessarily CSCs (see Nguyen et al. 2012 and the references therein). Considering CSCs as the apex of the hierarchy, they can undergo either symmetric or asymmetric divisions to replenish the CSC pool and to generate progenitor cells with limited proliferative potential and low tumorigenic potential. Typically, an early progenitor will divide into later (more mature) progenitors, undergoing only several rounds of self-renewing cell division before terminally differentiating. Emerging evidence supports the conclusion that early generations of progenitor cells share the same biomarker. Thus, we assume that the cell surface protein expression (we use the general notation CD+) is shared by CSCs and the first three generations of progenitor cells (Turner and Kohandel 2010, 2012), Figure 1.

Thus, we also define a $TCP_{CD+}$ as the probability of eliminating or controlling biomarker positive (CD+) cells.

A feature of the CSC hypothesis is its unidirectional nature. However, recent studies have supported the existence of considerable plasticity between the non-CSC and CSC populations, suggesting bidirectional conversions between these two compartments (Marjanovic et al. 2013). Such dedifferentiation may arise due to stochastic acquisition of genetic or epigenetic mutations in genes promoting the CSC-like state (for example, MBI1). Experimental studies have also shown that the reverse process can occur through the epithelial to mesenchymal transition (Mani et al. 2008, Chaffer et al. 2011). In this paper, we assume that dedifferentiation can be ignored during radiation treatment and consider a solely unidirectional hierarchy for the CSC hypothesis; future works in this area may include an analysis of how the rate of dedifferentiation impacts the TCP.

On the theoretical side, several models have been developed to study the TCP (see Zaider and Hanin 2001, for a review). A simplified model for the TCP can be computed based on binomial statistics, whereby we may define a success to be a cell death, and then the tumour control probability is defined to be the probability that there are $n_0$ successes, where $n_0$ is the total clonogenic cell population. This model neglects to include cell proliferation between fractions, as well as stochastic effects. A second model that has been studied extensively is based on Poisson statistics, used to approximate the stochastic process of radiation-induced cell killing. In this model, deterministic differential equations are formulated that account for cell growth and death due to both natural causes, as well as radiation, and then the probability of tumour control is given by a Poisson distribution, whose mean is the solution to the deterministic equation. A third model that has been formulated that describes the stochastic effects of radiation-induced cell death with great accuracy is the Zaider-Minerbo TCP model (Zaider and Minerbo 2000), in which the stochastic processes underlying cell birth and death are formalized, and used to define a master equation. Then, via a generating function approach, this master equation is transformed into a partial differential equation, which can subsequently be solved for the TCP. Recent works have concentrated on accounting for cell cycle effects on the potency of radiotherapy, by accounting for the extra radioresistance conferred by the presence of quiescent (non-active) tumour cells (Maler and Lutscher 2010, Dhawan et al. 2013).

Despite all of the extensions to the TCP that can be found in the literature, the primary argument leveled against the TCP, that it is not a measurable quantity during therapy, still holds. Because tumour control is only achieved when the entire clonogenic cell population has been eliminated, in order to experimentally verify this, it would be necessary to examine every remaining tumour cell for clonogenicity. The extension to the TCP that is presented here accounts for the presence of cancer stem cells, and because (barring effects of dedifferentiation during radiation) the eradication of cancer stem cells effectively implies tumour control, the $TCP_S$ is defined as the control of the cancer stem cell population only. In addition, due to the lack of a perfect biomarker, we define a second variant of the TCP, called $TCP_{CD+}$ defined only as the control of biomarker-positive cells. Moreover, the model used to describe stem cells in this work is a hierarchical one, which has not been previously considered in the literature.

We use stochastic methods, combined with analytical and computational techniques, to calculate $TCP_S$ and $TCP_{CD+}$. We show the relationship between these two variants of the TCP under different scenarios. Thus, we essentially depict the relationship between a measurable quantity and a theoretical quantity, and show that the proposed substitute measurable quantity is, generally, an effective surrogate for the theoretical TCP.

This novel formulation will be applicable as a measure of treatment success, in situations, for instance, where a biopsy of tumour cells is performed, and using cell detection protocols, the level of control of the biomarker positive cells is determined. From this, the level of overall tumour control can be inferred, and the radiation therapy adjusted to reach the therapeutic target computed by the theoretical TCP. In this way, using the measurable quantity of the control of the biomarker positive cells, the theoretical TCP can be estimated, and the goals of therapy can be accomplished with greater efficacy.

Additionally, for tumours with small numbers of cells, such as micro-metastases or those grown in vitro, the formulation of the TCP derived in this work will be of great benefit. In these cases, stochastic effects dominate, and thus, the TCP becomes a significant quantity in determining radiation dosing. Again, using data about the tumour in the form of the level of control of the biomarker positive cells at various time points along the course of the treatment, the theoretical TCP can be inferred, thus guiding the therapeutic protocol.

The practical application of the TCP, in a clinical setting, centres on its use in predicting treatment outcomes, comparing different treatment schedules. When used in conjunction with an appropriate model for the normal tissue complication probability (NTCP), the TCP can be used to determine an optimal radiation dose.

**Materials and methods**

The mathematical details concerning the full derivation and proof of the computation of both the theoretical and measurable TCP can be found in the Supplementary Information document. Here we present a short description of these techniques.

With the goal of developing a fully stochastic model for tumour growth and treatment by radiation, we first define a hierarchy of the critical cell populations within a tumour that are central for further analysis in a manner similar to Turner and Kohandel (2010). We consider three populations, stem $S$, progenitor $P$, and mature $M$, cells. Fundamentally, stem cells differentiate into progenitor cells, which differentiate into mature cells, or $S \to P \to M$. However, we note that while stem cells have the capacity for unlimited division, progenitor cells divide only a limited number of times, and mature cells do not divide. Thus, we assume that a progenitor cell divides exactly $N$ times before finally differentiating into a mature cell $M$. That is, we have the modified hierarchy $S \to P_1 \to \cdots \to P_N \to M$. Additionally, we note that any type of cell may be killed by the radiation, and we assume that it occurs with rate $\Gamma_i$, for $i = s, p, m$, representing stem, progenitor, and mature cells, respectively. Now, for the model, the following division and apoptosis pathways are active, with the rates for each type of division as follows (note that $r_1 + r_2 + r_3 = 1$ and $i = 1, \cdots, N$):

$$\begin{cases} S \to S + S & : \rho_s r_1 \\ S \to S + P_1 & : \rho_s r_2 \\ S \to P_1 + P_1 & : \rho_s r_3 \\ S \to 0 & : \Gamma_s \end{cases}, \quad \begin{cases} P_i \to P_{i+1} + P_{i+1} & : \rho_{p_i} \\ P_N \to M + M & : \rho_{p_i} \\ P_i \to 0_1 & : \Gamma_p \\ M \to 0 & : \Gamma_m \end{cases}$$

Where $\rho_s$ and $\rho_{p_i}$ refer to the birth rates for stem and progenitor cells, respectively, and $r_1$, $r_2$ and $r_3$ refer to the probabilities of each type of stem cell division. For the purposes of this model, as well as the subsequent analysis, we assume that cell deaths are independent of one another.

In order to compute the TCP, we must first define the joint probability function for the system. That is, we must define the probability that the system contains a given number of each type of cell at the time $t$. We make the assumption that at the initial time $t_0$, the number of each type of cell is known and these values are denoted $n_{0S}, n_{0P_1}, ..., n_{0P_N}, n_M$, for stem cells, each of the generations of progenitor cells, and mature cells, respectively.

We then proceed by defining a set of functions that give the probability for each possible combination of cells in each class, at any time. Using these functions, we derive a set of Master equations defined as the derivatives of these functions with respect to time. This information is used to isolate the function giving the probability of having eradicated all stem cells at a given time, which we can solve for analytically, thereby deriving the theoretical TCP. Using an analogous procedure, we obtain a differential equation defining the function representing the TCP for control of the biomarker positive cells (i.e. $S, P_1, P_2, ..., P_k$). This differential equation can be solved numerically by a novel method based on the method of characteristics, which is a well-established differential equation solution method (Supplementary Information).

**Results**

The details of the model used to describe the radiation induced cell-killing (i.e. the hazard function) are defined in the Supplementary Information. The model is a modified version of the linear-quadratic (LQ) model, with the primary assumption that all cell death occurs directly in the interval during the irradiation treatment. Within this model, there are radiosensitivity parameters $\alpha$ and $\beta$ that change based on the cell or tissue type, since different cells (e.g. biomarker-positive cells vs. biomarker-negative cells) have been shown to have different radioresponses (Baumann et al. 2008). For instance, Bao et al. (2006) have observed that CD133+ cells exhibit greater radioresistance than CD133- cells in human glioblastomas. Based on these data, following Turner et al. (2009), we assume that there is a three-fold increase in the LQ radio-sensitivity parameters, $\alpha$ and $\beta$, for biomarker-negative cells as compared to biomarker-positive cells. This depicts the fact that not only is control of these radio-resistant biomarker-positive cells highly desirable, because it theoretically provides control of stem cells, but also that accomplishing it would take a similar amount of radiation as control of all tumour cells, because of this difference in the radio-sensitivities. Figure 2 provides a graphic visualization of this difference in radiosensitivities as it pertains to survival fractions, or the fraction of a given amount of cells that survive irradiation for a given dose.

To illustrate the differences between using a measurable quantity as a suitable representative of the TCP, the $TCP_S$ (theoretical TCP) and $TCP_{CD+}$ (measurable TCP) values were computed numerically (supplemental Information) for three distinct radiation schedules, and compared. In the lack of experimental data for the number of biomarker positive generations of progenitor cells, $N$, for each of the radiation treatment schedules, the $TCP_{CD+}$ was computed for distinct cases of 1, 2, or 3 biomarker positive generations of progenitor cells. Using a conventional treatment schedule (2 Gy/fr, 5 fr/wk, up to 60 Gy), from Powathil et al. (2007), we compare the $TCP_S$ and the three curves generated for $TCP_{CD+}$, with either 1, 2, or 3 generations of progenitor cells as biomarker-positive in Figure 3. The results of the simulations show that fundamentally, for realistic biological parameters, the difference between the $TCP_S$ and $TCP_{CD+}$ curves is not drastic, as the curves reach a significant probability all near the same time point (Figure 3). That is, if the $TCP_{CD+}$ was used as a substitute for the $TCP_S$, tumour control would clinically correspond to 1-2 extra days of radiotherapy or 2-4 Gy of additional radiation, under the more conservative $TCP_{CD+}$ value, depending on the radiation schedule in question. These results are encouraging, and suggest that using the $TCP_{CD+}$ as a substitute for the theoretical value is indeed possible, and the extra cost of tumor control using a more conservative estimate is marginal.

Furthermore, three treatment protocols were compared directly: conventional, hyper-fractionated, and accelerated hyper-fractionated, taken from Powathil et al. (2007). The radiation schedules were as follows: conventional (scheme 1): 2 Gy/fr, 5 fr/wk, up to 60 Gy; hyper-fractionated (scheme 2): 1.2 Gy/fr, 2 fr/d, 5 d/wk, up to 60 Gy; accelerated hyper-fractionated (scheme 3): 1.5 Gy/fr, 2 fr/d, 5 d/wk, up to 60 Gy. The results of these simulations are depicted in Figures 4 and 5, it can be inferred that the $TCP_S$ curves obtained are qualitatively consistent with the $TCP_{CD+}$ curves, further suggesting that the measurable quantity is sufficient to act as a clinical substitute for the theoretical TCP.

It is also important to note that in the simulations considered, the effect of setting $\alpha_S = \alpha_P$, giving $\Gamma_S = \Gamma_P$, which gives a characteristic biomarker-positive cell death rate, is an increased separation of the TCP curves obtained for increasing numbers of progenitor cells in the biomarker-positive compartment. If this assumption is not made, then it could be assumed that stem cells are less radiosensitive than progenitor cells, thereby giving the result that the TCP obtained depends very heavily on this population, so looking at a biomarker-based TCP would be nearly identical to looking at the theoretical TCP, because the effect of the additional generations of early progenitor cells is so negligible.

**Discussion**

By describing a tumor model consisting of a unidirectional hierarchy of cancer stem cell proliferation into progenitor cells, and subsequently into mature cells within a tumor (Figure 1), and modeling the evolution of this system as a stochastic process, two quantities were generated and subsequently analyzed. Namely, the $TCP_S$ and the $TCP_{CD+}$, representing the probability of eradicating all CSCs and the probability of eradicating all biomarker-positive cells, respectively, were generated from our model via mathematical analysis and numerical computation. These quantities were then computed and compared for three different radiation therapy treatment schedules. The general conclusion is that the $TCP_{CD+}$, representing a measurable quantity, can reliably estimate the $TCP_S$, suggesting that using the $TCP_{CD+}$ as a clinical substitute for the theoretical $TCP_S$ is indeed very feasible.

The results obtained herein also provide a possible explanation in the context of certain contradictory experimental results obtained. Specifically, it was reported by McCord et al. (2009) that CD133+ glioblastoma cells isolated from two different neurosphere cultures did not display consistent radioresponse behaviour relative to corresponding CD133- cells (McCord et al. 2009). The obtained survival fraction curves showed that biomarker positive cells from one neurosphere culture were more radioresistant than the corresponding biomarker negative cells from the same tumour, but biomarker positive cells from a different tumour were found to be approximately the same radiosensitivity as biomarker negative cells from that tumour. Based on the results of this experiment, we hypothesize that the observed that the survival fraction (and corresponding TCP) varies, depending upon the proportion of stem cells contained within the biomarker positive compartment. From our perspective, this is explained since the biomarker positive cells are not necessarily a homogeneous population of cancer stem cells, but may also include generations of progenitor cells with a different radiosensitivity, so that the survival fraction curve for the biomarker positive cells can lie anywhere between the two survival fraction curves for a homogeneous population of stem cells only and a homogeneous population of progenitor cells (although the shape of the curve may not necessarily be the same). Thus, this observation serves as a possible explanation of the results that biomarker positive cells from different tumour cell lines do not necessarily give the same survival fraction curve because they do not necessarily consist entirely of the same proportion of stem cells, even through all cells in both samples are biomarker positive. In fact, the survival curve obtained can be rationalized as a form of a weighted (non-arithmetic) average of radiosensitivities of the two tumour cell subpopulations that can be classified as biomarker-positive.

While the results obtained in this work represent a novel computational approach to calculate the TCP, and add a measurable, experimentally quantifiable aspect to it (namely the $TCP_{CD+}$), an unaddressed limitation of the TCP is that it neglects to account for spatial effects within the tumour microenvironment that have been shown to contribute greatly to its radioresponse. One specific future direction in exploring this avenue of research would be to extend the presented model to include the effects of the oxygen enhancement ratio (Wouters and Brown 1997, Dasu et al. 2005, Powathil et al. 2011). Because the presence of oxygen is critical in radiobiology due to its role in the formation of free radicals, the rate of radiation-induced cell killing is highly dependent upon it. That is, within a clinical context, the OER would account for the effects of a hypoxic microenvironment and the changes in radiosensitivity of tumour cells. Using this data in conjunction with the TCP model would allow for the TCP to take into account spatial effects, by adjusting for local oxygen availability and distribution. The extension in this direction is primarily motivated by current research suggesting that the tumour microenvironment plays a key role in determining the properties of stem cells, and their corresponding radioresponse.

Another avenue for this research would be to combine the results presented here with previously studied compartment methods, such as the active-quiescent model (Maler and Lutscher 2010, Dhawan et al. 2013), to take full advantage of the radiobiological data. Doing so would allow for an even more accurate, and possibly clinically applicable formulation of the TCP, since factors not taken into account within the presented model, such as cell phase within the cell cycle, could be accounted for.

Finally, an alternative, experimental research direction is to quantify the effects of the radiotherapy, specifically by using the tools presented above to analyze the TCP by observable means. Doing so would require the generation of survival curves, and from this curve fitting to obtain radiosensitivity parameters. Using the results of such experiments would enable clinicians to utilize more clinically relevant parameters, and thereby improve the applicability of the results presented, ultimately improving patient outcomes (Baumann et al. 2009).

**Acknowledgements**

We thank K. Kaveh for useful discussions.

**References**

M. Al-Hajj, M. S. Wicha, A. Benito-Hernandez, S. J. Mor- rison and M. F. Clarke, Prospective identification of tumorigenic breast cancer cells, Proceedings of the National Academy of Sciences 100: 3983– 3988 (2003).

S. Bao, Q. Wu, R. E. McLendon, Y. Hao, Q. Shi, A. B. Hjelmeland, M. W. Dewhirst, D. D. Bigner and J. N. Rich, Glioma stem cells promote radioresistance by preferential activation of the DNA damage response, Nature 444: 756–760 (2006).

M. Baumann, M. Krause and R. Hill, Exploring the role of cancer stem cells in radioresistance, Nat Rev Cancer 8: 545–554 (2008).

M. Baumann, M. Krause, H. Thames, K. Trott, D. Zips, Cancer stem cells and radiotherapy, Int J Radiat Biol 85(5): 391-402 (2009).

D. Bonnet and J. E. Dick, Human acute myeloid leukemia is orga- nized as a hierarchy that originates from a primitive hematopoietic cell, Nature Medicine 3: 730–737 (1997).

C. L Chaffer, I. Brueckmann, C. Scheel, A. J. Kaestli, P. A. Wiggins, L. O. Rodrigues, M. Brooks, F. Reinhardt, Y. Su, K. Polyak, et al, Normal and neoplastic nonstem cells can spontaneously convert to a stem-like state, Proceedings of the National Academy of Sciences 108: 7950–7955 (2011).

A. Dasu, I. Toma-Dasu and M. Karlsson, The effects of hypoxia on the theoretical modelling of tumour control probability, Acta Oncologica 44: 563–571 (2005).

A. Dhawan, K. Kaveh, M. Kohandel and S. Sivaloganathan, Stochastic model for tumor control probability: Effects of cell cycle and (a)symmetric proliferation, Submitted to Phys. Med. Bio. (2013).

A. Maler and F. Lutscher, Cell-cycle times and the tumour control probability, Math. Med. Bio. 27: 313–342 (2010).

S. A. Mani, W. Guo, M. Liao, E. N. Eaton, A. Ayyanan, A. Y. Zhou, M. Brooks, F. Reinhard, C. C. Zhang, M. Shipitsin, et al, The epithelial-mesenchymal transition generates cells with properties of stem cells, Cell 133: 704–715 (2008).

N. D. Marjanovic, R. A. Weinberg and C. L. Chaffer, Cell plasticity and heterogeneity in cancer, Clinical Chemistry 59: 168–179 (2013).

A. M. McCord, M. Jamal, E. S. Williams, K. Camphausen and P. J. Tofilon. CD133+ glioblastoma stem-like cells are radiosensitive with a defective dna damage response compared with established cell lines, Clinical Cancer Research 15: 5145–5153 (2009).

T. R. Munro and C. W. Gilbert, The relation between tumour lethal doses and the radiosensitivity of tumour cells, British Journal of Radiology 34: 246–251 (1961).

L. V Nguyen, R. Vanner, P. Dirks and C. J. Eaves, Cancer stem cells: An evolving concept, Nature Reviews Cancer 12: 133–143 (2012).

G. Powathil, M. Kohandel, S. Sivaloganathan, A. Oza and M.Milosevic, Modeling the spatial mathematical modeling of brain tumors: Effects of radiotherapy and chemotherapy, Phys Med Biol. 52: 3291-306 (2007).

G. Powathil, M. Kohandel, M.Milosevic and S. Sivaloganathan, Modeling the spatial distribution of chronic tumor hypoxia: Implications for experimental and clinical studies, Comp. Math. Meth. Med., Article ID 410602 (2011).

S. K. Singh, I. D. Clarke, M. Terasaki, V. E. Bonn, C. Hawkins, J. Squire and P. B. Dirks, Identification of a cancer stem cell in human brain tumors. Cancer Res. 63: 5821–5828 (2003).

S. K. Singh, C. Hawkins, I. D. Clarke, J. A. Squire, J. Bayani, T. Hide, R. M. Henkelman, M. D. Cusimano and P. B. Dirks, Identification of human brain tumour initiating cells, Nature 432: 396– 401 (2004).

C. Turner, A. R. Stinchcombe, M. Kohandel, S. Singh and S. Sivaloganathan, Characterization of brain cancer stem cells: A mathematical approach, Cell Proliferation 42: 529-540 (2009).

C. Turner and M. Kohandel, Investigating the link between epithelial–mesenchymal transition and the cancer Stem cell phenotype: A mathematical approach, Journal of Theoretical Biology 265: 329-335 (2010).

C. Turner and M. Kohandel, Quantitative approaches to cancer stem cells and epithelial-mesenchymal transition, Seminars in Cancer Biology 22: 374-8 (2012).

B. G. Wouters and J. M. Brown, Cells at intermediate oxygen levels can be more important than the hypoxic fraction in determining tumor response to fractionated radiotherapy, Radiation Research 147: 541–550 (1997).

M. Zaider and G. N. Minerbo, Tumour control probability: A formulation applicable to any temporal protocol of dose delivery, Phys. Med. Bio. 45: 279-293 (2000).

M. Zaider and L. Hanin, Tumor control probability in radiation treatment, Med. Phys. 38: 574-83 (2011).

**List of Figure Captions**

Figure 1: The images, adapted from Singh et al. (2003), show that CD133+ tumor cells can proliferate in culture as non-adherent spheres, whereas CD133- tumor cells are not able to proliferate and form spheres. We assume that CSCs and early progenitors (from non-CSC compartment) share the CD133 biomarker. Here $S$, $P_i$ ($i = 1,...,N$), and $M$ denote stem, progenitors and mature cells, respectively.

Figure 2: A graph showing a comparison of the survival fraction, as a function of a single dose for biomarker-positive and biomarker-negative cells, assuming a three-fold increase in the radio-sensitivity parameters $\alpha$ and $\beta$ for biomarker-negative cells as compared to positive cells.

Figure 3: A panel showing a comparison for the measurable TCP curves when $k = 3$, for scheme 1 (conventional therapy), scheme 2 (hyper-fractionated therapy), and scheme 3 (accelerated hyper-fractionated therapy), noting that there is qualitative agreement with the theoretical TCP curves. The right graph is for TCP as a function of time in days, and the left is a for TCP as a function of dose of radiaton administered.

Figure 4: A panel showing a comparison for the theoretical TCP curves, for scheme 1 (conventional therapy), scheme 2 (hyperfractionated therapy), and scheme 3 (accelerated hyperfractionated therapy), noting that there is qualitative agreement with the measurable TCP curves. The right graph is for TCP as a function of time in days, and the left is a for TCP as a function of dose of radiaton administered.

Figure 5: A panel showing a comparison for the measurable and theoretical TCP curves for $k = 0,1,2,3$, for the conventional therapy (scheme 1), noting that as $k$ is increased, a slight shift occurs in the TCP curve. The right graph is for TCP as a function of time in days, and the left is a for TCP as a function of dose of radiaton administered.

**List of Figures**

**Figure 1**

CSCs    Non-CSCs

$S \rightarrow P_1 \rightarrow P_2 \rightarrow P_3 \rightarrow \cdots \rightarrow P_{N-1} \rightarrow P_N \rightarrow M$

CD133$^+$         CD133$^{--}$

**Figure 2**

Survival Fraction S(D) for Positive and Negative Cells

— Biomarker-positive Cells
— Biomarker-negative Cells

Dose (Gy)

**Figure 3**

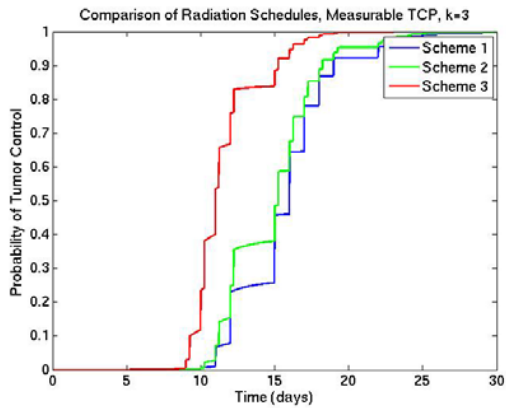 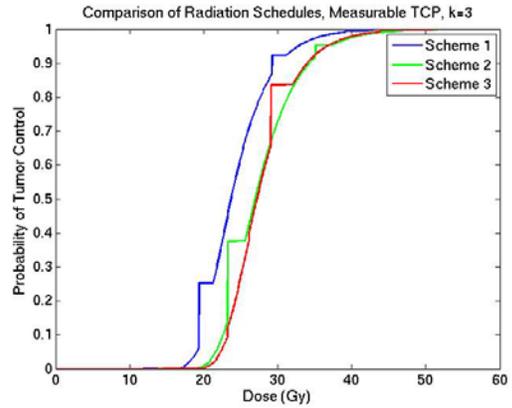

**Figure 4**

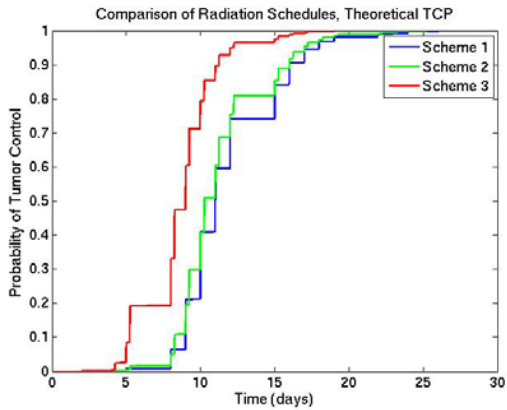 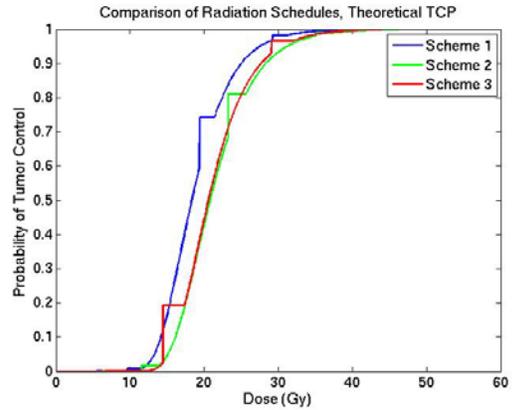

**Figure 5**

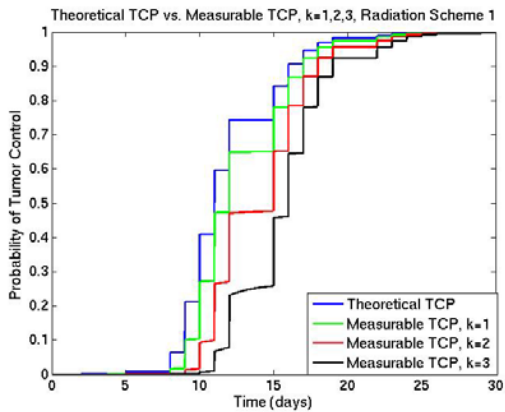 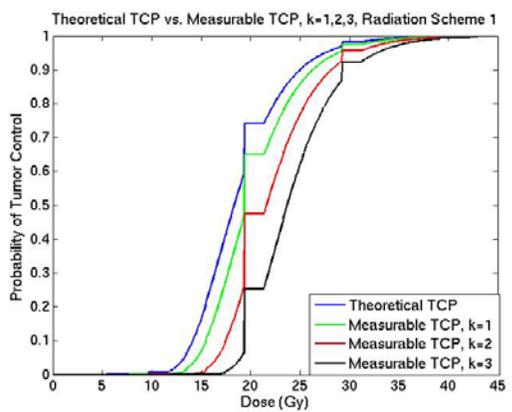

# Supplementary Information

## 1 Mathematical Model

With the goal of developing a fully stochastic model for tumor growth and treatment by radiation, we first define a hierarchy of the heterogenous cell populations within a tumor that are central for further analysis. Following Refs. [1, 2], we consider three populations, stem $S$, progenitor $P$, and mature $M$ cells. Fundamentally, stem cells differentiate into progenitor cells, which differentiate into mature cells, or $S \to P \to M$. However, we note that while stem cells have the capacity for unlimited division, progenitor cells divide only a limited number of times, and mature cells do not divide. Thus, we assume that a progenitor cell divides exactly $N$ times before finally differentiating into a mature cell $M$. That is, we have the modified hierarchy $S \to P_1 \to \ldots \to P_N \to M$ [1, 2]. Additionally, we note that any type of cell may be killed due to radiation, and we assume that it occurs with rate $\Gamma_i$, for $i = s, p, m$, representing stem, progenitor, and mature cells, respectively. The model includes the following division pathways (note that $r_1 + r_2 + r_3 = 1$, and $i = 1, \ldots, N$):

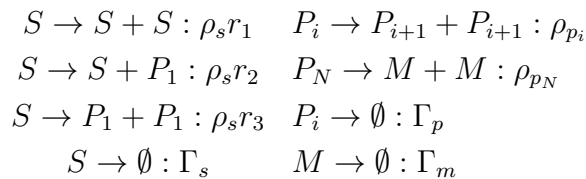

Here $\rho_s$ and $\rho_{p_i}$ refer to the proliferation rates for stem and progenitor cells, respectively, and $r_1, r_2, r_3$ refer to the probabilities of each type of stem cell division (symmetric self-renewal, asymmetric self-renewal, or symmetric commitment). For the purposes of this model, as well as the subsequent analysis, we assume that the cell deaths are independent of one another.

In order to compute the TCP, we first define the joint probability function for the system. That is, we define the probability that the system contains a given number of each type of cell at the time $t$ [3]. We make the assumption that at the initial time $t_0$, the number of each type of cell is known and these values are denoted $n_S^0, n_{P_1}^0, \ldots, n_{P_N}^0, n_M^0$, for stem cells, each of the generations of progenitor cells, and mature cells, respectively. We denote this probability function by $p_{n_S, n_{P_1}, n_{P_2}, \ldots, n_{P_N}, n_M}(t)$. The corresponding master equation [3] is then given by:

$$\begin{aligned}
\frac{dp}{dt} &= p_{n_S-1}\rho_s r_1(n_S - 1) + \rho_s r_2 n_S p_{n_{P_1}-1} \\
&+ \rho_s r_3(n_S + 1)p_{n_S+1,n_{P_1}-2} + \Gamma_s(n_S + 1)p_{n_S+1} \\
&- (\rho_s + \Gamma_s)n_S p \\
&+ \sum_{i=1}^{N-1} \rho_{P_i}(n_{P_i} + 1)p_{n_{P_i}+1,n_{P_{i+1}}-2} \\
&+ \sum_{i=1}^{N-1} (n_{P_i} + 1)p_{n_{P_i}+1}\Gamma_p - (\rho_{P_i} + \Gamma_p)n_{P_i} p \\
&+ \rho_{P_N}(n_{P_N} + 1)p_{n_{P_N}+1,n_M-2} - \rho_{P_N} n_{P_N} p \\
&+ \Gamma_m(n_M + 1)p_{n_M+1} - \Gamma_m n_M p \;.
\end{aligned} \quad (1)$$

We have omitted indices of $p(t)$ that remain unchanged for brevity. The initial condition is given by ($\delta_{i,j}$ is the Kronecker delta function):

$$p_{n_S,n_{P_1},n_{P_2},\ldots,n_{P_N},n_M}(t_0) = \delta_{n_S,n_S^0}\delta_{n_{P_1},n_{P_1}^0}\ldots\delta_{n_{P_N},n_{P_N}^0}\delta_{n_M,n_M^0} \;. \quad (2)$$

From this, we can obtain master equations for the probability functions for the number of stem cells only, as well as for the number of marker-positive cells ($S$, $P_1$, $P_2$, ..., $P_k$). Denoting the probability function for the number of stem cells, as $u_{n_S}(t)$, we observe that, by definition:

$$u_{n_S}(t) = \sum_{n_{P_1},\ldots,n_{P_N},n_M \geq 0} p_{n_S,n_{P_1},n_{P_2},\ldots,n_{P_N},n_M}(t)$$

Thus we obtain

$$\begin{aligned}
\frac{du_{n_S}}{dt} &= u_{n_S-1}\rho_s r_1(n_S - 1) \\
&- u_{n_S} n_S(\rho_s r_1 + \rho_s r_3 + \Gamma_s) \\
&+ u_{n_S+1}(\rho_s r_3 + \Gamma_s)(n_S + 1) \;,
\end{aligned} \quad (3)$$

with the initial condition $u_{n_S}(t_0) = \delta_{n_S,n_S^0}$. Following Van Kampen [3], and Zaider and Minerbo [5], we solve for $u_{n_S}(t)$ analytically by introducing the generating function $U(s,t) = \sum_{i=0}^{\infty} u_i(t)s^i$. We then obtain the following partial differential equation(PDE) for $U(s,t)$:

$$\frac{\partial U}{\partial t} - \frac{\partial U}{\partial s}(s - 1)(\rho_s r_1 s - \rho_s r_3 - \Gamma_s) = 0 \;. \quad (4)$$

This PDE can be solved via the method of characteristics [4], along with the initial condition $U(s,0) = s^{n_S^0}$. Assuming that all parameters are constant, we obtain

$$U(s,t) = \left[\frac{(s-1)(\rho_s r_3 + \Gamma_s)e^{(\rho_s(r_1-r_3)-\Gamma_s)t} - \rho_s r_1 s + \rho_s r_3 + \Gamma_s}{(s-1)(\rho_s r_1)e^{(\rho_s(r_1-r_3)-\Gamma_s)t} - \rho_s r_1 s + \rho_s r_3 + \Gamma_s}\right]^{n_S^0} \;. \quad (5)$$

From this, we observe that the probability that there are no stem cells remaining at time $t$ is given by

$$U(0,t) = \left[\frac{(\rho_s r_3 + \Gamma_s)e^{(\rho_s(r_1-r_3)-\Gamma_s)t} - \rho_s r_3 - \Gamma_s}{(\rho_s r_1)e^{(\rho_s(r_1-r_3)-\Gamma_s)t} - \rho_s r_3 - \Gamma_s}\right]^{n_S^0}. \quad (6)$$

We thus define this quantity to be the theoretical TCP, denoted $\text{TCP}_S(t)$.

Using the same technique, we can obtain the probability function for the number of marker positive cells, denoted $v_{n_S, n_{P_1}, \ldots, n_{P_k}}(t)$. We have:

$$v_{n_S, n_{P_1}, \ldots, n_{P_k}}(t) = \sum_{n_{P_{k+1}}, \ldots, n_{P_N}, n_M \geq 0} p_{n_S, n_{P_1}, n_{P_2}, \ldots, n_{P_N}, n_M}(t). \quad (7)$$

This gives (where again we have omitted the indices which remain unchanged, for brevity):

$$\begin{aligned}
\frac{dv}{dt} =\ & v_{n_S-1}\rho_s r_1 (n_S - 1) + \rho_s r_2 n_S v_{n_{P_1}-1} \\
& + \rho_s r_3 (n_S + 1) v_{n_S+1, n_{P_1}-2} \\
& + \Gamma_s (n_S + 1) v_{n_S+1} - (\rho_s + \Gamma_s) n_S v \\
& + \sum_{i=1}^{k-1} \rho_{P_i} (n_{P_i} + 1) v_{n_{P_i}+1, n_{P_{i+1}}-2} \\
& + \sum_{i=1}^{k-1} (n_{P_i} + 1) v_{n_{P_i}+1} \Gamma_p - (\rho_{P_i} + \Gamma_p) n_{P_i} v \\
& + (\Gamma_p + \rho_{P_k})(v_{n_{P_k}+1}(n_{P_k} + 1) - v n_{P_k}).
\end{aligned} \quad (8)$$

The initial condition is $v_{n_S, n_{P_1}, \ldots, n_{P_k}}(t_0) = \delta_{n_S, n_S^0} \delta_{n_{P_1}, n_{P_1}^0} \ldots \delta_{n_{P_k}, n_{P_k}^0}$. To solve for $v(t)$, similar to the previous case, we introduce the generating function,

$$V(x_S, x_1, \ldots, x_k, t) = \sum_{i_S, i_1, \ldots, i_k \geq 0} v_{i_S, i_1, \ldots, i_k}(t) x_S^{i_S} x_1^{i_1} \ldots x_k^{i_k}. \quad (9)$$

We can obtain the following PDE for $V(x_S, x_1, \ldots, x_k, t)$:

$$\begin{aligned}
& \frac{\partial V}{\partial x_S}\left(x_S^2 \rho_s r_1 - (\rho_s + \Gamma_s - \rho_s r_2 x_1) x_S\right) \\
& + \frac{\partial V}{\partial x_S}\left(\rho_s r_3 x_1^2 + \Gamma_s\right) \\
& + \left(\sum_{l=1}^{k-1} \frac{\partial V}{\partial x_l}\left(\rho_{P_l} x_{l+1}^2 - (\rho_{P_l} + \Gamma_p) x_l + \Gamma_p\right)\right) \\
& + \frac{\partial V}{\partial x_k}(\Gamma_p + \rho_{P_k})(1 - x_k) = \frac{\partial V}{\partial t},
\end{aligned} \quad (10)$$

with the initial condition $V(x_S, x_1, \ldots, x_k, 0) = x_S^{n_S^0} x_1^{n_{P_1}^0} \ldots x_k^{n_{P_k}^0}$. This PDE can be numerically solved via a modified method of characteristics, which is outlined in Dhawan et al. [6].

## 2 Results

The model defining the effects of radiation on the cell populations is essentially the same as that defined in [7]. Other model parameters were taken from [8], except for $\beta_s, \beta_p$, which were calculated from data collected in [7], and $\omega$, which was estimated specifically for glioblastoma treatment. The parameters used are as follows: $\rho_s = 0.6931$ (1/day), $\rho_p = \rho_{p_i} = 0.6931$ (1/day), $r_1 = 0.15$, $r_2 = 0.7$, $r_3 = 0.15$, $\alpha_s = 0.2\,\text{Gy}^{-1}$, $\beta_s = 0.02\,\text{Gy}^{-2}$, $\alpha_p = 0.2\,\text{Gy}^{-1}$, $\beta_p = 0.02\,\text{Gy}^{-2}$, $n_S^0 = 100, n_{P_1}^0 = 100$, $n_{P_2}^0 = 100$, $n_{P_3}^0 = 100$, $\omega = 15\,\text{min}$. The hazard function, a function describing radiation-induced cell death, is based on the linear-quadratic (LQ) model for cell survival for each dose fraction [5]. Thus, the hazard function for an individual dose fraction of radiation, with dose $d$ and irradiation period $\omega$, takes value:

$$f_j(t,d) = \frac{(\alpha_j + 2\beta_j d) \cdot d}{\omega}, \qquad (11)$$

during the period of irradiation ($0 \leq t \leq \omega$), and for all $t > \omega$ or $t < 0$, is identically 0 (with $j = s, p$). Using this, we define the overall function describing the rate of cell death for the entire treatment, consisting of all dose fractions as:

$$\Gamma_j(t) = \sum_i f_j(t - t_i, d_i), \quad j = s, p. \qquad (12)$$

In the above expressions, $\alpha_s, \alpha_p, \beta_s, \beta_p$ are parameters describing the radiosensitivities of stem ($s$) and progenitor ($p$) cells in the LQ model, $t_i$ is the time of the $i^\text{th}$ fraction, $d_i$ is the dose administered on the $i^\text{th}$ fraction, and the sum is over all fractions administered.

## References

[1] C. Turner and M. Kohandel, Investigating the Link between epithelial-mesenchymal transition and the cancer stem cell phenotype: A mathematical approach, Journal of Theoretical Biology 265: 329-335 (2010).

[2] C. Turner and M. Kohandel, Quantitative approaches to cancer stem cells and epithelial-mesenchymal transition, Seminars in Cancer Biology 22: 374-8 (2012).

[3] N. G. Van Kampen, Stochastic Processes in Physics and Chemistry, North-Holland Personal Library: Elsevier Science (2007).

[4] L. C. Evans, Partial Differential Equations, Graduate Studies in Mathematics: American Mathematical Society (1998).

[5] M. Zaider and G. N. Minerbo, Tumour control probability: A formulation applicable to any temporal protocol of dose delivery, Phys. Med. Bio. 45: 279-293 (2000).

[6] A. Dhawan, K. Kaveh, M. Kohandel and S. Sivaloganathan, Stochastic model for tumor control probability: Effects of cell cycle and (a)symmetric proliferation, Submitted to Phys. Med. Bio. (2013).

[7] G. Powathil, M. Kohandel, S. Sivaloganathan, A. Oza and M.Milosevic, Modeling the spatial mathematical modeling of brain tumors: Effects of radiotherapy and chemotherapy, Phys Med Biol. 52: 3291-306 (2007).

[8] C. Turner, A. R. Stinchcombe, M. Kohandel, S. Singh and S. Sivaloganathan, Characterization of brain cancer stem cells: A mathematical approach, Cell Proliferation 42: 529-540 (2009).